\documentclass[epj]{svjour}
\begin{document}
\title{'Universal' FitzGerald Contractions}
\author{Merab Gogberashvili}
\institute{Andronikashvili Institute of Physics, 6 Tamarashvili
Street, Tbilisi 0177, Georgia \and Javakhishvili State University,
3 Chavchavadze Avenue, Tbilisi 0128, Georgia \and 
California State University, 2345 E. San Ramon Avenue M/S MH37,
Fresno, CA 93740-8031, USA}
\date{Received: \today}
\abstract{The model of a universe with a preferred frame, which nevertheless shares the main properties with traditional special and general relativity theories, is considered. We adopt Mach's interpretation of inertia and show that the energy balance equation, which includes the Machian energy of gravitational interactions with the universe, can imitate standard relativistic formulas. 
 }

\maketitle


\section{Introduction}

The special theory of relativity is traditionally based on the relativity principle together with Einstein's postulate of the constancy of the speed of light \cite{Ein}. However, relativity is not a branch of electromagnetism and can be developed without any reference to light. It is known that the Lorentz transformations can be derived using only the relativity principle, supplemented by the assumptions of homogeneity, isotropy and smoothness \cite{I-postulate}. The existence of a universal speed follows as a consequence of the relativity principle. The FitzGerald contraction factor in the Lorentz transformations can be expressed more generally as
\cite{Fitz},
\begin{equation} \label{gamma}
\gamma = \sqrt{1 + \frac{v^2}{\Phi}}~,
\end{equation}
which instead of the speed of light contains some universal parameter $\Phi < 0$, which should be determined from experiment. Indeed, Lorentz-type transformations appear even in the equation for sound waves, where the characteristic constant is the sound velocity in the media \cite{Lamb}. Experiments actually show that the numerical value of the universal parameter $\Phi$ in (\ref{gamma}) is equal to the square of the light speed in vacuum. Due to this fact the speed of light became an essential part of the Minkowski space-time formalism.

Besides of the relaxation of the speed of light invariance postulate there are some other attempts to modify traditional special relativity by models taken from both ends of the energy scale.

In the IR one deals with the wide class of gedanken experiments (originated from the Einstein, Podolsky and Rosen paradox), which, by breaking Bell's inequalities, manifest a strong non-locality in quantum mechanics \cite{Nonlocal}. The possibility of non-causal information transfer between distant parts of the system is in contradiction with Einstein's locality principle.

The problems of orthodox relativity in the UV are connected with the question about the meaning of the Planck quantities in cosmology and quantum gravity. It is known that the ranges of the values of all physical observables are limited by the Planck quantities and these limits should not depend on the chosen inertial reference frame \cite{Planck}.

Another non-clear subject of Einstein's relativity is the concept of empty space, which apparently is in conflict with the quantum theory \cite{Quant}. Quantum field theory is well defined in the presence of vacuum fluctuations while the background space-time still remains flat. However, according to the Einstein equations the non-zero vacuum energy, in general divergent, should induce curved geometry and such divergences promise serious difficulties. So far there is no experimental evidence for the vacuum energy to couple to gravity, while it is believed that the vacuum fluctuations are real as experimentally verified by the Casimir effect. In some models the energy of vacuum fluctuations, including the divergent parts, possesses the non-zero gravitational mass demanded by the equivalence principle \cite{Mil}. However, this does not give the final solution since by allowing the vacuum energy to be large one creates another difficulty -- the cosmological constant problem. 

In this paper we try to reformulate relativity arguments to keep local Lorentz invariance in the IR and at the same time incorporate the cosmological preferred frame, which is observed. 

The main question arising in a model with a preferred frame is how to explain the relativity principle, which is the foundation of Newton's and Einstein's formalisms. In these theories the concept of mass and the behavior of 'free' bodies play a key role in the definition of geometry, since trajectories of free bodies are independent of their masses and compositions and it is possible to regard them as properties of space-time. In this way one arrives at the kinematics of the Galilean and Riemann spaces. As expected in a model with a preferred frame the mass-independent formulation should exists only in an Aristotelian space-time, in which the natural state of motion is rest. Note that Aristotelian space-time probably is the underlining geometry of the de Broglie-Bohm formulation of the standard quantum theory \cite{deBr}. 

We claim that in the model with a preferred frame the relativity principle (the first postulate of special relativity) can be retrieved as a consequence of the isotropy and homogeneity of matter distribution in the universe (the cosmological principle) if we adopt Mach's interpretation of inertia \cite{Mach}. Indeed, in the homogeneous universe there exists a class of privileged observers, having constant velocities with respect to the preferred frame, for which the universe appears spherically symmetric (the standard definition of inertial frames \cite{Wei}), and thus the total Machian forces on them are zero. We also want to replace the assumption of a fundamental velocity (the second postulate of special relativity) by introducing a 'universal' gravitational potential. Then our model can exhibit local Lorentz invariance in spite of the existence of a preferred frame, i.e. non-local gravitational interactions with the universe can imitate standard relativistic formulas in the IR. This approach has several novel features: the relativistic effects pertain only to matter fields, while gravitation singles out a cosmological frame and thus remains non-relativistic; Minkowski space-time corresponds to the case of the isotropic and homogeneous universe and not to the empty vacuum with zero energy as in general relativity; the speed of light and Newton's constant both become effective quantities and depend on the matter content of the universe.


\section{Energy balance condition}

Note that in many cases it is inconvenient to use parameters such as lengths and masses to describe relativistic systems. Some examples of difficulties of using the traditional formalism are: non-clear 'resolutions' of the twin paradox, existence of different kind of non-physical horizons and appearance of velocity dependent mass parameters. In our opinion, it is better to use conserved quantities, such as the energy of a system, as the basic variable and try to define other parameters through it. Energy is one of the most disputed topics in general relativity, however, it is well defined in the Newtonian theory. So, for the simplicity, we shall use the Newtonian concept of energy -- the ability to do work. Also we allow the existence of a preferred frame in our model. Then the energy is the conserved quantity associated with the cosmic time translations for an observer at rest. 

Using the energy additivity as in the Newtonian realm we write the energy balance equation for a particle as:
\begin{equation}\label{W}
E = E_0 + T + U~.
\end{equation}
Here $E_0$ represents the energy of the non-local gravitational interactions of the particle with the universe, $T$ is its kinetic energy with respect to the preferred frame and $U$ corresponds to all local interactions. The new ingredients in (\ref{W}), even within the Newtonian approach, are the introduction of the non-local gravitational energy, $E_0$, and apparently the absence of the relativity principle (because $T$ depends on the velocity with respect to the preferred frame). However, aspects of relativity can be explained within the model with a preferred frame if we use Mach's principle. Then the relativity principle recovers because in the homogeneous universe (when $U = 0$) the value of $E_0$ should be the same for all inertial observers (having any $T$). 

Let's make some idealizations and replace the distant universe by a spherical shell of the order of the horizon. Then the Newtonian potential theory can be used to estimate $E_0$ -- the energy of the non-local gravitational interaction due to the distant masses. Homogeneity and isotropy inside of the shell directly follow from the cosmological principle for these distances. The shell acts similar to a gravitational Faraday cage inside which there exists the following 'universal' gravitational potential
\begin{equation} \label{Phi}
\Phi = - \frac{M_g G}{R} = const ~,
\end{equation}
where $G$ is Newton's constant. The total gravitational mass of the universe $M_g$ (which includes all dark components as well) can be found from the critical density condition, which is valid even in Newtonian cosmology \cite{Bar}. The radius of the universe $R$, or the radius of the causal sphere of any object, can be estimated using the experimental values of the Hubble constant and the speed of light. Then the numerical value of the gravitational potential of the universe on a particle appears to be equal to the square of the speed of light \cite{Gog,Sia},
\begin{equation} \label{phi=c}
\Phi = - c^2~.
\end{equation}
It seems that this relation is correct for all stages of the universe expansion since Planck times \cite{Ken}, because
\begin{equation} \label{Pl}
\frac{G M_P}{l_P} = c^2~,
\end{equation}
where $l_P$ and $M_P$ are Planck's length and mass respectively. In fact, (\ref{Pl}) serves as the definition of the Planck mass in terms of Newton's constant, $G \sim 1/M_P^2$ (in units where $c = 1$).

We adopt the formulation of Mach's principle which states that the origin of a particle's inertia lies in the interactions of this particle with all other gravitating particles inside its causal sphere \cite{Gog}. By inertia of a particle we mean the ratio of its total energy with the 'universal' gravitational potential,
\begin{equation} \label{m}
m = - \frac{E}{\Phi}~,
\end{equation}
in which using (\ref{phi=c}) we recover Einstein's famous energy-inertia relation. 

Note that already in 1925 E. Schr\"{o}dinger considered a Machian model similar to ours \cite{Sch}. He also modeled distant masses as a uniform spherical shell and identified the gravitational potential energy of the shell with the particle's kinetic energy. When he considered the radius of the shell to be of the order of the size of the Milky Way, he concluded that the inertia of particles must primarily be due to matter external to our galaxy. Indeed, as we mentioned above, to have the realistic value (\ref{phi=c}) for the 'universal' potential one should consider the shell having the size of the Hubble sphere \cite{Gog,Sia}.


\section{Hierarchy problem}

The model of a 'free' particle at rest with respect to the universe, i.e. when in (\ref{W}) we assume
\begin{equation}
T = U = 0~,
\end{equation}
was considered in our previous paper \cite{Gog}. The particle's energy (\ref{W}) in this case coincides with the energy of its non-local gravitational interactions with the universe,
\begin{equation} \label{E_0}
E = E_0 = - m_g \Phi ~,
\end{equation}
where $m_g$ is some coupling constant that can be interpreted as the particle's gravitational mass and $\Phi$ is the 'universal' potential on the particle. The definition of inertia (\ref{m}) in this case reduce to the equivalence principle, 
\begin{equation} \label{m=E}
m_0 = m_g ~. 
\end{equation}
However, the introduction of the gravitational mass in (\ref{E_0}) is ambiguous and in general (\ref{m}) and (\ref{E_0}), instead of (\ref{m=E}), give:
\begin{equation} \label{Phi=c}
 m_0 = m_g~\frac{M_gG}{c^2 R}~.  
\end{equation}
This relation can be used to address the hierarchy problem in particle physics \cite{Gog}, i.e. it can be used to reduce the Plank scale $M_P \sim 1/\sqrt G \sim 10^{19} ~GeV$ to the Higgs scale $M_H \sim 1 ~TeV$. In fact, (\ref{Phi=c}) is unchanged if together with the assumption that inertial and gravitational masses are not equivalent but are proportional to one another,
\begin{equation} \label{m=Nm}
m_g = \sqrt{N} m_0~,
\end{equation}
($N \sim 10^{40}$ is the amount of gravitating particles within the horizon), we introduce the new gravitational constant
\begin{equation} \label{g=NG}
g = N G \sim 1/M_H^2~.
\end{equation}
The assumption (\ref{m=Nm}) does not mean a violation of the weak equivalence principle, which is well tested \cite{EP}, since for inertial observers in a homogeneous and isotropic universe $N$ is conserved and is still the same for all bodies, regardless of their composition.


\section{Inertial frames}

For an inertial particle, i.e. when $U = 0$, the energy balance condition (\ref{W}) has the form:
\begin{equation}\label{E+T}
E = E_0 + T ~.
\end{equation}
In this case the parameter of inertia (\ref{m}) becomes velocity depended \cite{Inertia},
\begin{equation} \label{m=E/c}
m = \frac{E}{c^2} = m_0 + \frac{T}{c^2}~,
\end{equation}
while the mass $m_0 \sim m_g$, which characterized the particle itself, is the frame-independent quantity.

As it was mentioned above the shell-universe model leads to the standard definition of inertial frames \cite{Wei} because $\Phi$ in the homogeneous universe is constant. Then the 'universal' force on a uniformly moving particle is zero and there is positional invariance inside the shell. Furthermore any inertial observer finds himself at the center of his own uniform causal shell and the universe appears spherically symmetric for him (relativity principle). 

Since the Machian energy, $E_0 = - m_g \Phi$, is the same for all inertial observers, we can replace the special relativity condition of intervals invariance by the condition of the equality of energies,
\begin{equation} \label{E=E}
E_0 = E - T = const~.
\end{equation}
Then from (\ref{E+T}) and (\ref{m=E/c}) follows: 
\begin{equation} \label{dE}
c^2 dm - d T = 0~.
\end{equation}
Using Hamilton's equations and the definition of momentum $p^i = mv^i$ ($i = 1,2,3$) from this relation we have \cite{So-Pi}:
\begin{eqnarray} \label{int}
c^2 dm - d T = c^2 dm - v^i dp_i = \nonumber \\
= \left( c^2 - v^2 \right) dm - \frac m2  dv^2 = \\
= m \left( c^2 - v^2 \right) d \left( \ln m\sqrt{1 - 
\frac{v^2}{c^2}}\right) = 0~. \nonumber
\end{eqnarray}
Thus we find that the quantity
\begin{equation} \label{m0}
m_0 = m\sqrt{1 - \frac{v^2}{c^2}}~,
\end{equation}
which we identify with the inertial mass (\ref{m=E}), is constant. So in our model the well known 'relativistic' effect of the particle's inertia growth depending on its velocity has a dynamical nature and is a consequence of the energy balance condition (\ref{E+T}) and not of the Lorentz transformations. 

We can state that all other 'relativistic' effects, such as length contraction and time dilation, have also dynamical nature. Indeed, examining the formula of the linear momentum,
\begin{equation} \label{p}
p^i = m v^i = \frac{m_0v^i}{\sqrt{1 - v^2/c^2}}~,
\end{equation}
we notice that the FitzGerald correction factor (\ref{gamma}) can be attached to $dt$ rather than to $m_0$ and the notion of 'proper time' can be introduced
\begin{equation} \label{tau=t}
d\tau = dt\sqrt{1 - \frac{v^2}{c^2}}~.
\end{equation}
In this way the velocity-dependence of the definition of inertia (\ref{m=E/c}) can be described by the kinematics of Minkowski space-time, where the model gains the simplest form. So we conclude that the contraction (\ref{tau=t}) is a dynamical effect and is the result of the dilation of the rest energy of moving systems in the background gravitational field of the universe. Also from (\ref{m=E/c}) and (\ref{E+T}) one can obtain that the quantity
\begin{equation}
E_0^2 = E^2 - c^2 p^2~,
\end{equation}
where $E_0 = m_0c^2$, is time independent, i.e. 
\begin{equation}
dE_0 / dt = 0~. 
\end{equation}

At the end of this section let us note that one can use the non-local Machian energy of a particle, $E_0$, to define its action integral,
\begin{equation} \label{S}
S = \int_{t_1}^{t_2} dt~ m_g \Phi ~.
\end{equation}
If the starting and the ending times $t_1$ and $t_2$ correspond to uniform velocities, the universality of Machian energy for inertial frames (our version of the relativity principle) leads to the principle of least action,
\begin{equation}
\delta S = 0~.
\end{equation}
Also (\ref{S}), according to (\ref{E+T}), gives the usual expression of the action integral for a 'free' particle (which in our interpretation interacts with the whole universe),
\begin{equation} \label{S1}
S = - \int dt~ m_0c^2 \sqrt{1 - \frac{v^2}{c^2}} = - m_0c \int d\tau~.
\end{equation}


\section{Non-inertial cases}

Now consider the general energy balance condition (\ref{W}). In the simplest case when only the weak local gravitational potential $\phi$ is present, i.e. $T=0$ and $U = m \phi$, we have:
\begin{equation} \label{Phi=phi}
E = E_0 +  m \phi~ = m_0 \Phi_{tot}~,
\end{equation}
where we supposed $m = m_0$ and introduced the notion of total gravitational potential,
\begin{equation}
\Phi_{tot} = - \Phi + \phi = c^2\left(1 + \frac{\phi}{c^2}\right)~.
\end{equation}
This relation can be understood as the modification of light velocity by the local gravitational potential $\phi$, or as the light diffraction near gravitating body. Alternatively (\ref{Phi=phi}) can be rewritten using the time dilation factor, 
\begin{equation} \label{tau}
d\tau = dt\sqrt{1 + \frac{\phi}{c^2}}~.
\end{equation}
Usually it is assumed that the relativistic slowing of moving clocks and the slowing of clocks in lower gravitational potentials are different effects. In our model both terms in the general contraction factor (the combination of (\ref{tau=t}) and (\ref{tau})),
\begin{equation} \label{phi}
d\tau = dt\sqrt{1 + \frac{\phi}{c^2} - \frac{v^2}{c^2}}~,
\end{equation}
have gravitational nature and correspond to the interactions of a particle with a local gravitating source and with the whole universe, respectively.

Now let us consider (\ref{W}) with an arbitrary $U$. Since for the accelerated particle the universe no longer looks like a spherical shell, the 'universal' potential $\Phi$ is not a constant, and instead of (\ref{dE}) we should write 
\begin{equation}\label{dW}
dE = - m_g d\Phi + dT + dU~,
\end{equation}
where $m_g d \Phi = m_g W_i dx^i$ and $dU = F_i dx^i$ are the works done by the 'universal' gravitational field,
\begin{equation}
W_i = \partial_i \Phi~,
\end{equation}
and by the local forces,
\begin{equation}
F_i = m a_i
\end{equation} 
($a_i$ denotes acceleration), acting on a particle through the distance $dx^i$. The consequence of the relation (\ref{dW}), along with (\ref{dE}), is the principle of dynamical equilibrium \cite{Assis}: The balanced gravitational pull of the universe creates an opposing 'inertia' of massive objects in responding to local disturbances,
\begin{equation} \label{W=a}
m_g W_i - m a_i = 0~.
\end{equation}
This equation shows that local inertial force can be derived from the gravitational interactions with the universe, 'explains' the weak equivalence principle and is a quantitative implementation of Mach's hypothesis about the origin of inertia \cite{Mach}. 

The two equations, (\ref{m0}) and (\ref{W=a}) (which follow from the energy balance condition (\ref{dW})), guarantee the local Lorentz invariance and the stability of the 'gravitationally entangled' universe when its parts are in arbitrary motion. These equations also allow us to express the acceleration by the 'universal' gravitational potential,
\begin{equation}
a_i = \partial_i \Phi \sqrt{1 - \frac{v^2}{c^2}}~.
\end{equation}

Note that the presented model is compatible with the three hypotheses that are enough to describe gravity as the curvature of space-time. These three hypotheses are the weak equivalence principle, local Lorentz invariance and position invariance \cite{Will}. The weak equivalence principle is represented in our model by the relation (\ref{m=Nm}). Local Lorentz invariance and position invariance follow from the energy balance condition (\ref{W}). Then, since matter fields couple universally to a gravitational field, one can consider the metric as a property of space-time itself rather than as a field over space-time and switch to the geometrical formulation in four dimensions.


\section{Conclusions}

To summarize in this paper we show that the Machian model can lead to the main features of the special and general relativity theories in spite of the existence of a fundamental frame. The universal cosmic parameter with dimensions of velocity square (needed for the FitzGerald contraction factor (\ref{gamma})), the formulas of special relativity and the justification of the 4-dimensional geometrical formulation, naturally appear in this framework. Additionally, the model opens the way to solve the hierarchy problem in particle physics and provides insight towards formulating a non-relativistic quantum theory of gravity.

Note that our approach have some similarities with the Ho\~{r}ava-Lifshitz model of gravity \cite{Horava}, which also becomes relativistic only in the IR. This model also assumes that the effective speed of light and Newton's constant emerge from the deeply non-relativistic theory at UV. In contrast to Ho\~rava's model our approach still lacks a complete formulation that will be done in forthcoming papers. 


\acknowledgement{{\bf Acknowledgment:} Author would like to acknowledge the support of a 2008-2009 Fulbright Fellowship.


\end{document}